\UseRawInputEncoding

\documentclass[aps,prx,longbibliography,a4paper,notitlepage,reprint,nofootinbib
]{revtex4-2}

\usepackage{graphicx}
\usepackage{braket}
\usepackage{color}
\usepackage{soul}
\usepackage{float}
\usepackage{physics, amsmath, amssymb, color, siunitx}

\usepackage{hyperref}
\hypersetup{colorlinks=true,
    linkcolor=blue,
    filecolor=blue, 
    citecolor=blue,
    urlcolor=blue,
}

 \usepackage[normalem]{ulem}

\makeatletter
\newcommand{\fmarki}{*}
\newcommand{\fmarkii}{\ensuremath{\dagger}}
\def\@fnsymbol#1{{\ifcase#1\or \fmarki\or \fmarkii \else\@ctrerr\fi}}
\makeatother
\renewcommand{\fmarki}{$\dagger$}
\renewcommand{\fmarkii}{$\ddagger$}
\newcommand{\had}{\hat{a}^{\dagger}}
\newcommand{\ha}{\hat{a}}

\begin{document}
\title{\textbf{Electro-optic conversion of itinerant Fock states}}

\author{T. Werner$^\star$}%
\email{Contact author: thomas.werner@ist.ac.at}
\affiliation{
 Institute of Science and Technology Austria (ISTA), Am Campus 1, 3400 Klosterneuburg, Austria}
 \author{E. Riyazi$^\star$}
\affiliation{
 Institute of Science and Technology Austria (ISTA), Am Campus 1, 3400 Klosterneuburg, Austria}
\author{S. Hawaldar}
\affiliation{
 Institute of Science and Technology Austria (ISTA), Am Campus 1, 3400 Klosterneuburg, Austria}
\author{R. Sahu}
\affiliation{
 Institute of Science and Technology Austria (ISTA), Am Campus 1, 3400 Klosterneuburg, Austria}
 \author{G. Arnold}
\affiliation{
 Institute of Science and Technology Austria (ISTA), Am Campus 1, 3400 Klosterneuburg, Austria}
 \author{P. Falthansl-Scheinecker}
\affiliation{
 Institute of Science and Technology Austria (ISTA), Am Campus 1, 3400 Klosterneuburg, Austria}
 \author{J. A. S\'anchez-Naranjo}
\affiliation{
 Institute of Science and Technology Austria (ISTA), Am Campus 1, 3400 Klosterneuburg, Austria}
 \author{D. Loi}
\affiliation{
 Institute of Science and Technology Austria (ISTA), Am Campus 1, 3400 Klosterneuburg, Austria}
 \author{L. N. Kapoor}
\affiliation{
 Institute of Science and Technology Austria (ISTA), Am Campus 1, 3400 Klosterneuburg, Austria}
  \author{M. Zemlicka}
\affiliation{
 Institute of Science and Technology Austria (ISTA), Am Campus 1, 3400 Klosterneuburg, Austria}
 \author{L. Qiu$^\diamond$}
\affiliation{
 Institute of Science and Technology Austria (ISTA), Am Campus 1, 3400 Klosterneuburg, Austria}
\author{A. Militaru}
\affiliation{
 Institute of Science and Technology Austria (ISTA), Am Campus 1, 3400 Klosterneuburg, Austria}
\author{J. M. Fink}%
\email{Contact author: jfink@ist.ac.at}
\affiliation{
 Institute of Science and Technology Austria (ISTA), Am Campus 1, 3400 Klosterneuburg, Austria}

\date{\today}

\begin{abstract}

Superconducting qubits are a leading candidate for utility-scale quantum computing due to their fast gate speeds and steadily decreasing error rates. The requirement for millikelvin operating temperatures, however, creates a significant scaling bottleneck. Modular architectures using optical fiber links could bridge separate cryogenic nodes, but superconducting circuits do not have coherent optical transitions and microwave-to-optical conversion has not been shown for any non-classical photon state.
In this work, we demonstrate the on-demand generation 
and tomographic reconstruction 
of itinerant single microwave photons at 8.9~GHz from a superconducting qubit.
We upconvert this non-Gaussian state with a transducer added noise below 0.012 quanta and count the converted telecom photons at 193.4~THz with a signal-to-noise ratio of up to 5.1$\pm$1.1. We characterize the trade-offs between throughput and noise, and establish a viable path toward heralded entanglement distribution and gate teleportation. Looking ahead, these results 
empower existing superconducting devices to take a key role in distributed quantum technologies and
heterogeneous quantum systems.
\end{abstract}

\maketitle

\def\thefootnote{$\star$}
\footnotetext{These authors contributed equally to this work.}
\def\thefootnote{\arabic{footnote}}

\def\thefootnote{$\diamond$}
\footnotetext{Present address: School of Physics, Zhejiang University, Hangzhou 310058, China.}
\def\thefootnote{\arabic{footnote}}

The relentless demand for computational power continues to drive innovation across both fundamental science and engineering. Quantum technologies, in particular, are gaining significant momentum, promising transformative advances in simulation~\cite{daley2022}, cryptography~\cite{pirandola2015}, and sensing~\cite{degen2017}. Among the various platforms, superconducting qubits have emerged as a front-runner due to their compatibility with established nanofabrication techniques, which enable precise parameter control and scalability~\cite{krantz2019}. Their high-speed operation---characterized by nanosecond gate~\cite{googleaiquantum2020}, readout~\cite{walter2017} and reset times~\cite{magnard2018}---coupled with strong coupling to microwave fields~\cite{blais2021}, makes them an ideal candidate for large-scale quantum processors.

However, the very microwave-frequency operation that enables high-speed control also presents a fundamental scaling bottleneck. At room temperature, 
a $10$~GHz microwave mode is swamped by thermal noise, with an average occupancy of $\bar{n}_\textrm{th} \approx 700$ photons. Consequently, superconducting circuits must remain confined to millikelvin environments, severely limiting their application in long-range quantum communication and networking~\cite{ giovannetti2001,duan2001,
yin2017,wehner2018}. In contrast, telecom-wavelength light ($f_o \approx 193$~THz) is effectively noise-free at room temperature, and---together with ultra-low loss and cost--- makes optical fiber the ideal medium for distributing quantum information 
~\cite{marcikic2004,ursin2007,krutyanskiy2023a}.

Bridging the microwave and optical regimes requires coherent and quantum limited microwave-to-optical conversion, a process essential for building future quantum networks~\cite{barzanjeh2012,han2021,krastanov2021}. Current experimental efforts to develop these transducers span a wide array of physical systems, including electro-optic (EO)~\cite{arnold2025a,warner2025,zhou2025}, electro- and piezo-optomechanic~\cite{forsch2020,zhao2025,urmey2025,vanthiel2025,meesala2024}, atomic vapor~\cite{kumar2023,borowka2024}, and spin ensemble~\cite{xie2025} systems. To date, these devices have enabled groundbreaking demonstrations such as qubit readout and control~\cite{delaney2022,vanthiel2025,arnold2025a}, microwave photon addition~\cite{jiang2023}, qubit-to-photon transduction~\cite{mirhosseini2020}, and the conversion of classical coherent states~\cite{fan2018,forsch2020,tu2022,kumar2023,borowka2024,blesin2024,sahu2022}. Furthermore, recent progress has extended this to continuous and discrete variable entanglement~\cite{sahu2023, meesala2024a}. 

Despite these advances, the conversion of an itinerant non-Gaussian state remains a critical, outstanding milestone. In this work, we use a transmon qubit encased in a three-dimensional aluminum resonator to generate a single microwave photon in the resonator. After this single photon leaks out from the resonator and propagates through the output waveguide, we direct it towards an EO transducer and upconvert it to the optical domain. We verify the non-classical nature of the generated single photon by performing state tomography in the microwave domain, and we quantify the signal-to-noise ratio of the conversion process using single-photon detection in the optical domain. 

Importantly, by upconverting an {\it itinerant} microwave photon, we ensure that the conversion process does not affect the superconducting qubit used to generate the photon. Our initial demonstration therefore does not only show that a non-classical state can be converted to the optical domain with high signal-to-noise ratio. It also overcomes a major hurdle in quantum interconnects for superconducting circuits, paving the way for multi-platform entanglement distribution~\cite{cirac1997} and for integrating superconducting nodes into existing photonics quantum networks~\cite{krastanov2021}. \color{black}

\section{Experimental Implementation}

\begin{figure}[t]
\includegraphics{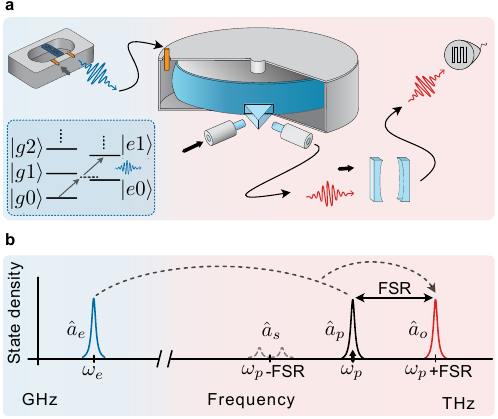}
\caption{\textbf{Schematic view of the experimental setup.} \textbf{a}, The blue shaded area shows a transmon qubit (teal) in a microwave cavity and the dispersively coupled system's energy level diagram. The cavity has a weakly coupled input port used to feed a two-photon drive to the transmon, and a stronger coupled output port used to collect the generated single photon (blue). A coaxial cable connects the cavity and the EO transducer (light blue disk). The optical pump (black arrow) couples out of an optical fiber and enters the transducer. The generated single infrared photon (red) propagates towards a series of filter cavities. Those are represented by two mirrors which prevent the reflected pump (black arrow) from reaching the single photon detector (grey).
\textbf{b}, Sketch of the spectrum depicting the four electromagnetic modes involved in the upconversion process. The blue, grey, black, and red modes represent the microwave $\omega_e$, the hybridized Stokes $\hat{a}_s$, the pump $\hat{a}_p$ with frequency $\omega_p$ and the signal (anti-Stokes) mode $\hat{a}_o$, respectively. 
\label{fig:Setup}}
\end{figure}

The central component in our setup is an electro-optic (EO) transducer which we sketch in Fig.~\ref{fig:Setup}a. It is based on the Pockels effect~\cite{pockels1893} which, enhanced by an optical pump, can change the frequency of a microwave (MW) photon to an optical one or vice-versa. The EO interaction is described by the interaction Hamiltonian~\cite{tsang2010, hease2020a} \[ \hat{H} = \hbar g_0 (a_e a_p a^{\dagger}_o + a^{\dagger}_e a^{\dagger}_p a_o) \approx \hbar \sqrt{n_p} g_0 (a_e a^{\dagger}_o + a^{\dagger}_e a_o).\] It assumes a system that couples three resonant modes. The strongly occupied optical pump mode $a_p$ can be seen as a classical state with the real amplitude $\sqrt{n_p}$ with $n_p$ being the average photon number population inside the pump mode. Using a classical pump field results in a beam-splitter interaction between the MW cavity mode $a_e$ and the optical mode $a_o$ with the enhanced vacuum coupling rate $\sqrt{n_p}g_0$. The process is illustrated in Fig.~\ref{fig:Setup}b. The hybridization of the Stokes mode $a_s$ with a second orthogonally polarized mode breaks the frequency-matching condition that would lead to a two-mode squeezing interaction between the microwave mode and the Stokes optical mode~\cite{rueda2016}.

The physical EO device
consists of a 3D superconducting MW cavity with an embedded disk-shaped LiNbO$_3$ optical whispering gallery mode resonator (WGMR).
Analogous to our previous work in Ref.~\cite{hease2020a},
the transducer's MW cavity is frequency-tuneable. We match it's second resonance to the optical resonator's free spectral range $\textrm{FSR} = 8.9006~\textrm{GHz}$, which is fixed by its radius of 2.5~mm (see Appendix~\ref{appendix:device_parameters} for device parameters) to achieve both energy conservation and phase matching.
Electric field confinement and EO mode overlap is achieved by 
evaporating thin-film aluminum directly on the top and bottom surfaces of the 500~$\mu$m-thick single crystal LiNbO$_3$. This was proposed in Ref.~\cite{rueda2019} and differs from the edge-clamped design used in previous works \cite{hease2020a,sahu2022,sahu2023,qiu2023,arnold2025a}. We find that thin film electrodes together with center-clamping allows for lower losses, easier assembly and better reproducibility, which also enables multiple thermal cycles without optical degradation. 

The transducer and the qubit-cavity are placed at the millikelvin stage of a dilution refrigerator setup (see Appendix~\ref{appendix:SI_mw_setup}). Optical fibers guide light inside the cryostat, and via gradient index lenses and a diamond prism, it is coupled  into the WGMR (see Appendix~\ref{appendix:optical_setup}). We use a 200~ns optical pump pulse of $P_p =$ 1.22~mW at the disk-prism interface for all optical measurements in this work.

To avoid thermal heating and quasiparticle generation, we operate the transducer in the so-called regime of low cooperativity $C \approx 4.1\times10^{-4}$, where the rate of internal EO conversion is much lower than the dissipation in both resonators \cite{hease2020a}. We realize an internal EO conversion efficiency of $\eta_{\textrm{int}} \approx 1.6 \times10^{-3}$ and a total efficiency (including both in- and outcoupling losses) of $\eta_{\textrm{ext}} \approx 2.2\times10^{-4}$. 
We note that the power efficiency $\eta_{\textrm{ext}}/P_p$ is comparable to our previous results in the high cooperativity limit~\cite{sahu2022} even though the vacuum coupling strength $g_0/2\pi = 4.2~\textrm{Hz}$ is nine times smaller due to non-optimized microwave mode confinement in this first generation thin-film electrode device.

Besides the transducer, the second crucial component of our setup is a MW single photon (SP) source, in our case a transmon qubit~\cite{koch2007} in a 3D aluminum cavity (from now on called qubit-cavity). The qubit-cavity is connected to the EO transducer~\cite{rueda2019} via a coaxial cable from one of its ports. From another weakly-coupled port, we drive a blue sideband (BSB) transition between the qubit-cavity joint states $\ket{g,0}$ and $\ket{e,1}$ to generate single MW photons~\cite{kindel2016,campagne-ibarcq2018}. 
Specifically, when both the qubit and the resonator are in their ground state, the BSB pulse excites the qubit to the $\ket{e}$ state while occupying the cavity with Fock state $\ket{1}$. The qubit-cavity is frequency-tunable~\cite{arnold2025} and, when the qubit is in the excited state, it is matched to the optical FSR and the EO cavity resonance.

Another essential part of the setup is optical filtering. The transducer's output optical signal co-propagates with the reflected pump through the output optical fiber and is detuned from it in frequency by exactly one FSR. We filter this output using a chain of four cascaded Fabry-Perot cavities (55~MHz bandwidth), which are resonant with the signal frequency (43.7\% transmission) and strongly suppress the pump (170~dB suppression), see Appendix~\ref{appendix:optical_setup}. 

The final part of our setup is a photon-counting detector. We use a superconducting nano-wire SP detector (SNSPD) with a detection efficiency of $\approx 85\textrm{\%}$ to reconstruct the temporal profile of the Fock state after upconversion. Together with the signal photons we may also detect pump photons leaking through the signal cavity filters, Brillouin noise generated by the pump~\cite{yu2020}, and dark counts~\cite{yang2007}. We call these unwanted, uncorrelated detections optical noise (see Appendix~\ref{appendix:optical_noise}). In addition, the microwave cavity inside the transducer may be populated with thermal photons, which also get upconverted during the transduction process. The resulting single photon signal-to-noise ratio (SNR) is then determined by the combination of microwave losses before transduction, upconverted thermal photons (the transducers input referred added noise) and optical noise.

\section{Single photon Fock states}

By sending a MW pulse to the qubit-cavity at the qubit frequency, we can bring the qubit to the superposition state $\alpha\ket{g} + \beta\ket{e}$, with arbitrary, normalized $\alpha$ and $\beta$ ($\abs{\alpha}^2 + \abs{\beta}^2 = 1$). If this is followed by a BSB pulse, we implement the state $\ket{e} (\beta\ket{0} + \alpha\ket{1})$ inside the qubit-cavity. This state can then leak out into a propagating mode inside the output waveguide. In case $\alpha = \beta = 1/ \sqrt{2}$, we call the leaked state a half photon (HP) state. If $\beta = 0$ the qubit was in the $\ket{g}$ state and the BSB pulse produces a single photon (SP).

\begin{figure}[t]
\includegraphics{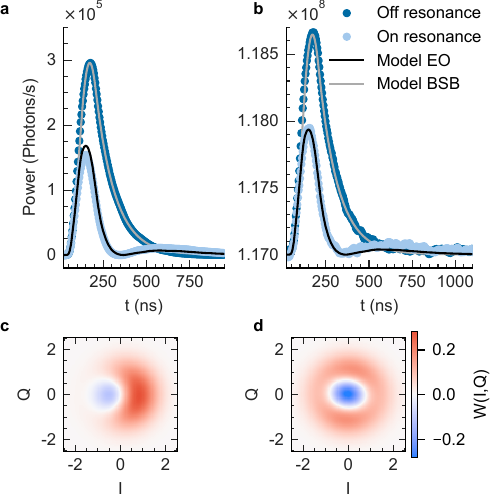}
\caption{\textbf{Microwave Fock state shapes and Wigner functions.} 
\textbf{a}, Time-resolved MW measurements of $\abs{\ev{\ha(t)}}^2$ for a prepared half photon state. 
\textbf{b}, Time-resolved MW measurements of $\ev{\had \ha (t)}$ for a prepared single photon state.
The light blue (dark blue) curves in both panels are for the qubit-cavity resonant (off-resonant) with the EO MW cavity. The power axes in both panels are calibrated via the measured noise offset in units of quanta in panel \textbf{b} and multiplied with the RBW. The grey solid lines are theory of the photon generation process and the black solid lines also take into account the resonant interaction with the EO MW cavity.
\textbf{c} (\textbf{d}), Wigner functions of the HP (SP) state calculated from the off-resonant experimental data. The fidelities for HP and SP states are 99.7\% and 97.6\%, respectively.
\label{fig:FockStates}}
\end{figure}

In order to characterize the state leaking out of the qubit-cavity after the BSB pulse, we collect the microwave radiation reflected off the EO resonator. We amplify this signal with a Josephson parametric amplifier (JPA) 
and measure it via heterodyne detection  with a 50 MHz resolution bandwidth (RBW), see Appendix~\ref{appendix:SI_mw_setup}. We distinguish two cases, one where the EO MW cavity is on resonance with the qubit-cavity, and the other where it is detuned by 13~MHz. The latter case enables us to analyze the emitted photon state without EO MW cavity interaction. 

We first conduct a coherent measurement of the HP to extract the amplitude $\ev{\ha(t)} = \ev{I(t) + iQ(t)}$, meaning we average the two field quadratures ($I,Q$) across all shots to extract the temporal envelope of the half photon~\cite{pechal2014}. Then, to obtain the resulting coherent power $\abs{\ev{\ha(t)}}^2=\ev{I(t)}^2 + \ev{Q(t)}^2$, we square and sum the two mean values for the quadratures, as shown in Fig.~\ref{fig:FockStates}a. It is possible to perform this measurement because the HP state has a fixed phase relation between the vacuum state and the SP Fock state leading to $\ev{a} \ne 0$. 

In contrast, for a SP state, there is no contribution of the $\ket{0}$ state to provide a phase reference $\ev{a}=0$.
Therefore, in Fig.~\ref{fig:FockStates}b, we implement an incoherent power measurement of the SP state and show the expectation value $\ev{\had \ha (t)} = \ev{I(t)^2 + Q(t)^2}$, meaning the quadratures are squared and then averaged, leading also to an integration of the incoherent background noise. A free scaling factor is fixed by setting this incoherent noise floor to $1.17 \times 10^8~\textrm{photons/s}$. This is the product of the RBW and the sum of the independently calibrated input referred added noise (1.84~photons/s/Hz) of the heterodyne detection obtained from a temperature controlled $50~\Omega$ termination in the output line and vacuum noise (see Appendix~\ref{appendix:SI_mw_losses_and_noise}). 

The off-resonant HP and SP shapes (dark blue data in Fig.~\ref{fig:FockStates}a and b) show a sharp rise followed by a decay with $1/\kappa \approx$ 100~ns caused by the qubit-cavity unloading. Limited BSB drive power, in combination with the fast decay time, causes the initial, finite rise which, in the ideal case, should be near-vertical~\cite{kindel2016}. We derive the corresponding theory (grey curve in Fig.~\ref{fig:FockStates}a) from a \verb +QuTiP+-based simulation of the qubit-cavity and the applied BSB pulse shape~\cite{johansson2013} (see Appendix~\ref{appendix:SI_mw_losses_and_noise}). 

When the qubit-cavity and the EO resonator are on resonance (light blue data in Fig.~\ref{fig:FockStates}a and b) we measure photon shapes with a smaller amplitude and faster decay due to the partial absorption in the EO resonator. In addition, we observe a small revival that is due to interference of the in-coupled and out-coupled signals. Theory taking into account the resonant response of the EO cavity (black curves in Fig.~\ref{fig:FockStates}a and b, see also Appendix~\ref{appendix:SI_mw_losses_and_noise}) is in excellent agreement with the time dependent data and demonstrates full understanding of the dynamics and nature of the generated single photon states.

Following Refs.~\cite{buzek1996,eichler2011}, we reconstruct the  Wigner function of the HP and SP states. First, we calculate a mode matching function (MMF) using the off-resonant curve from Fig.~\ref{fig:FockStates}b. Then we conduct a measurement comprising $10^{8}$ single shots. Using the MMF as integration weights, each measurement yields a point in the IQ-plane. We run this measurement interleaved for three different types of states, once generating a SP, once a HP, and once a vacuum state. Using the vacuum state for a reference measurement of the added noise, we can calculate the moments of the measured field's probability distributions and from there the Wigner functions. These functions, which show a clear negativity and follow the expected shapes of HP and SP states, are depicted in Fig.~\ref{fig:FockStates}c and d.
 
The SP state has a vanishing mean value $|\langle a \rangle| =$~0.03 (ideally 0) and a photon number of $\langle a^\dagger a \rangle =$~0.95 (ideally 1) because a small portion of vacuum is mixed in due to imperfect state preparation. Also for the HP state, $|\langle a \rangle| =$~0.51 and $\langle a^\dagger a \rangle = $~0.49, are close to the ideal value of 0.5. Furthermore, the fidelities for HP and SP are 99.7\% and 97.6\%, respectively.

\begin{figure*}[t]
\includegraphics[width=1\textwidth]{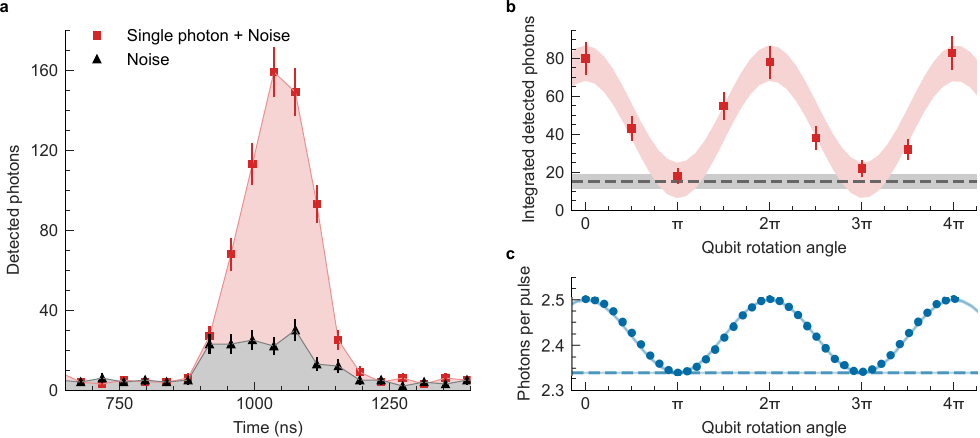}
\caption{\textbf{Optical single photon shape and single photon Rabi measurements.}  
\textbf{a}, Upconverted single MW photons measured in the optical domain. The red (black) curve represents detected photons vs. detection time when there's one (no) single MW photon being sent. The data are shown as absolute number of detected photons $N \pm \sqrt{N}$ within a 40~ns wide interval. The error bars represent one standard deviation. 
\textbf{b}, Single photon Rabi measurement in the optical (red) and \textbf{c}, the MW (blue) domain, show detected photons as a function of the  qubit rotation angle prepared by linearly sweeping the qubit drive amplitude. The data points are integrated over 200~ns for the optical and 800~ns for the MW measurement. The error bars for the MW measurement are smaller than its markers. The error bars for the optical measurements are calculated by assuming a Poissonian distribution.
We fit both data sets with a squared cosine, and show the 68\% confidence intervals of the fits as the shaded areas (not visible in the MW case). 
The dashed lines describe the minima of the respective functions and the errorbars represent one standard deviation.
\label{fig:RabiMeasurements}} 
\end{figure*}

\section{Single photon Rabi measurement}

We upconvert the on-resonant single MW photons by sending a synchronized optical pump pulse to the transducer. In our experiment, where the conversion bandwidth of the transducer (1.3~MHz) is comparable to the linewidth of the single photon source (1.4~MHz), both being much smaller than the optical linewidth (11.4~MHz), transduction happens through a ``load-and-convert'' scheme. 
First the itinerant MW photon loads into the EO MW mode. Only when the photon has close-to-maximally populated the MW mode, we load the optical pump mode to upconvert the MW photon. This 
scheme has the advantage of
converting only when the signal occupation is much greater than the residual thermal occupancy of the MW mode. 
In this work we realize a maximum single photon population of the EO MW mode of $N_{\textrm{cav,SP}} \approx 0.13$, for the SP itinerant state, which is due to mode mismatches, losses in the MW setup path and intrinsic losses in the two cavities (see Appendices~\ref{appendix:SI_mw_losses_and_noise} and~\ref{appendix:SI_mw_losses_and_SNR}).

We then filter, count and time-tag the upconverted optical photons with an SNSPD with a time 
resolution of 250~ps, allowing us to reconstruct the temporal profile in an averaged measurement.

In Fig.~\ref{fig:RabiMeasurements}a, we show results of an interleaved measurement where at every second trigger event a MW SP Fock state is sent to the transducer and in between no photon is generated. The conversion measurements with the incident vacuum state are used to estimate the noise that contaminates the SP signal. The red data in Fig.~\ref{fig:RabiMeasurements}a contain SP signal and noise counts. To obtain a valid SNR we therefore subtract the measured noise from the measured signal before dividing the result by the measured noise. 

We can clearly see that the signal surpasses the noise considerably in a 200~ns window when the optical pulse occupies the pump mode. This happens around 950~ns after the start of the detection window. If we restrict ourselves to this specific time span and integrate all detected signal (s = 470) and noise (n = 106) photons, we can calculate the SNR = 4.4~$\pm$~0.6. This measurement was run at a trigger rate of 1~kHz and around $6\times 10^7$ optical pulses were sent. To get a smooth curve describing the photon's shape, we show the sum of all detected photons in 40~ns time bins. Finally, we note that the SNR of the upconverted single infrared photons exceeds one by more than five (5.7) standard deviations.

We now use optical photon detection to observe the correlation between the upconverted signal photons and the qubit's initial state~\cite{houck2007}. Rotating the qubit around the x-axis on the Bloch sphere before generating the single MW photon and then upconverting it, allows us to run an optical single photon Rabi measurement~\cite{bertet2005,clarke2008}. We show this measurement in Fig.~\ref{fig:RabiMeasurements}b, where we sent $47 \times 10^6$ pump pulses to the transducer at a rate of 1~kHz. We perform Rabi oscillations up to an angle of $4\pi$. The measurement time for each of the ten points (nine for the distinct Rabi angles and one for noise) is 1.9~h.

Figure~\ref{fig:RabiMeasurements}c shows a single photon Rabi measurement in the MW domain (via heterodyne detection) for comparison. Here, we sent $41 \times 10^7$ BSB pulses at a rate of 100~kHz. The total measurement time is 4100~s. For this measurement, we detune the EO MW cavity, weight the microwave output signal with the MMF and then integrate, and average. The $\textrm{SNR} \approx 0.07$ is dominated by the input referred added noise of the heterodyne detection, the losses between 
the qubit-cavity and the point in the setup where this added noise is calibrated (4.7~dB), and the MMF used to integrate the signal, see Appendix~\ref{appendix:SI_mw_losses_and_SNR}). 

The optical domain has a clear advantage in SNR. This is because the single-photon detection of the SNSPD is insensitive to the vacuum noise that mixes in with the signal as a consequence of losses. The MW's advantage is the short measurement time due to faster repetition rates without thermal heating, and because we detect a higher portion of the signal due to lower losses. This is also reflected in the better statistics and the resulting smaller relative error of the averaged MW measurement.

\section{Noise characterization}
Signal-to-noise ratio, throughput, and input referred added noise are essential parameters for current transducers~\cite{zhao2025}. They give an idea of potential future applications and experiments that can be conducted with a device. Now that we have shown that our transducer has noise characteristics that enable quantum networking ($\textrm{SNR} \gg 0.5$) at reasonably low repetition rates (1~kHz), we examine its potential and limitations by reproducing the measurement shown in Fig.~\ref{fig:RabiMeasurements}a over a wide range of experimental repetition rates.

\begin{figure}[t]
\includegraphics{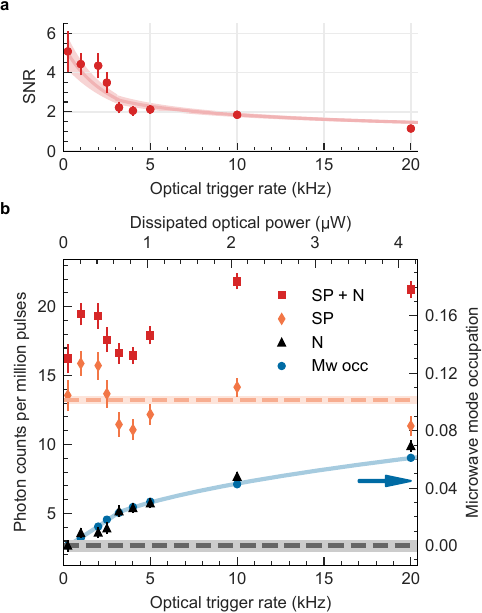}
\caption{\textbf{EO transducer's signal, noise, and SNR vs optical trigger rate.} 
\textbf{a}, SNR extracted from optical photon counting measurements presented in panel \textbf{b}. The time-resolved data for 1~kHz trigger rate are shown in Fig.~\ref{fig:RabiMeasurements}a. Error bars represent one standard deviation. The red line and its confidence interval (shaded region) are calculated from fitted noise (blue circles in panel \textbf{b}) and mean single photon detection efficiency (orange dashed line in panel \textbf{b}). 
\textbf{b}, Measured signal and noise counts per 1 million applied optical pulses. We show the average dissipated power on the top axis. The grey dashed line represents the optical noise measured in the absence of signal photons. The orange line represents the average detection probability for 1 million repetitions. The grey and orange shaded areas represent one standard deviation. The blue dots show the MW mode occupancy inferred from MW noise spectroscopy (right axis). The solid blue line consists of two power-law fits to the MW occupation. The first $\propto P_\textrm{diss}^{1.06}$
fits up to $\approx0.6\mu\textrm{W}$ dissipated power, the second $\propto P_\textrm{diss}^{0.43}$
from there~\cite{hease2020a}. 
The error bars represent one standard deviation of the mean. For the microwave measurements (blue circles) they are smaller than the marker size.
\label{fig:Noise}}
\end{figure}

Figure~\ref{fig:Noise}a shows the observed decrease of the single photon SNR from 5.1$\pm$1.1 at 250~Hz to 1.2$\pm$0.1 for a 20 kHz repetition rate. This is expected due to the increasing average dissipated optical power $P_\textrm{diss}$ leading to thermal population of the EO MW mode.
Like in the measurements shown in Fig.~\ref{fig:RabiMeasurements}b, we restrict ourselves to integrating over a specific region (200 ns) in the time-resolved measurement to achieve the highest values in SNR for the individual trigger rates. On average, we measured roughly $70 \times 10^6$ pump pulses for each rate. 

In Fig.~\ref{fig:Noise}b, we explicitly show the corresponding noise measurement results (black triangles), the measured single photon signal including noise (red squares), and the single photon signal (orange diamonds), which is calculated from the former two. 
Normalizing photon counts with the number of applied pulses allows us to compare signal and noise for different trigger rates independent of the absolute number of counts. The top axis shows the calculated dissipated optical power $P_\textrm{diss} = P_p (1-(1-2\Lambda^2\eta_o)^2) \,D$~\cite{hease2020a, arnold2025a}, where $\eta_o \approx 0.68$ is the optical coupling efficiency, $\Lambda \approx 0.66$ the optical mode overlap, and $D$ the duty cycle (trigger rate $\times$ optical pulse length). 

As expected for a constant pump power we find that the signal counts per pulse are roughly constant as a function of trigger rate and we extract a mean single photon detection efficiency of $\eta_{
\textrm{det}} = (1.32\pm0.03) \times 10^{-5}$. This is consistent with the transducer's conversion efficiency together with independently calibrated losses and a temporal mode mismatch of $\approx 0.5$, see Appendix~\ref{appendix:SI_mw_losses_and_noise}. We attribute the observed fluctuations around this mean value, described as the orange dashed horizontal line in Fig.~\ref{fig:Noise}b, to small changes of the MW cavity's parameters and possibly also in part to imperfect laser and filter locking.

We recognize two contributions to the measured noise. 
Optical noise manifests as a constant offset of $2.6\times 10^{-6}$ counts per pulse (dashed grey line), see details in Appendix \ref{appendix:optical_noise}. The second noise contribution is from upconverted thermal photons occupying the transducer's MW cavity. 

To confirm and quantify this we infer the thermal occupation of the MW cavity mode shown in Fig.~\ref{fig:Noise}b (blue circles, right axis) with continuous noise measurements using a MW spectrum analyzer based on an independent noise floor calibration. Similar to Ref.~\cite{hease2020a}, the measured MW noise occupancy first follows a linear (up to $0.6~\mu\textrm{W}$), then a square root rise as a function of $P_\textrm{diss}$, which we attribute to a temperature dependent change in thermal conductivity of the transducer. 

To relate this thermal occupation of the MW mode to the measured optical noise counts, 
we scale the MW noise occupancy axis with the mean efficiency of photon detection $\eta_{\textrm{det}}$, but now referred to the transducer's MW cavity occupancy $\eta_{\textrm{det,cav}} = \eta_{\textrm{det}} / N_{\textrm{cav,SP}}$. 
Also taking into account the optical noise offset we find excellent agreement in Fig.~\ref{fig:Noise}b (blue circles with fit and black triangles). 
This confirms that the trigger rate dependent reduction in SNR is due to optical absorption heating in the WGMR. The red line in  Fig.~\ref{fig:Noise}a shows the mean efficiency (orange line in panel b) divided by the scaled and offset fits to the MW occupation (blue line in panel b) and agrees with the data (red circles) very well.

We note that the transducer operates at very low thermal mode occupancies over the full range of measurement rates, i.e. $N_e$ between 0.005 (for 1 kHz) and 0.06 quanta (for 20 kHz) as confirmed from both, MW heterodyne and single photon counting measurements. For the coldest data obtained with a 250~Hz measurement rate we can not reliable extract $N_e$ due to a absence of a noise peak in the output spectrum over a reasonable integration time as well as the finite accuracy of the optical noise measurement. The input referred added noise of the transducer is then given as $N^{\textrm{up}}_{\textrm{add}}=N_e/\eta_e$, which ranges from 0.012 (for 1 kHz) to 0.12 quanta (for 20 kHz). 

These remarkably low numbers position this device clearly in the quantum limit with a positive quantum channel capacity, see Appendix \ref{appendix:throughput}. It also suggests that there is substantial room for SNR improvement without changing the transducer, e.g. with more narrow band filtering of optical noise or by reducing losses between the single photon source and the transducer.

\section{Conclusion and Outlook}

In this work, we generated non-classical itinerant microwave states and upconverted them to the optical domain, demonstrating direct correlations between prepared qubit states and optical photon counts. Rigorous noise analysis identifies optical fiber noise and upconverted thermal photons as the primary SNR constraints at low and high repetition rates, respectively.

Our results also establish a roadmap for future performance. The SNR determines the fidelity of entangled states~\cite{beukers2024} and as example we calculate an upper bound for the entanglement fidelity of a joint superconducting qubit and traveling optical photon state of 
$F_\textrm{Bell}=87.6\%$ 
(with the currently achieved maximum SNR) assuming a perfect MW Bell state preparation, placing us well in the quantum regime (see Appendix~\ref{appendix:Bell_fidelity} and~\ref{appendix:SI_optical_impact}). To improve this SNR further, we propose (i), reducing optical noise  by a factor of six using narrower signal filters, and (ii), by mitigating the loss between the qubit and EO cavity with superconducting cables, optimizing out-coupling of the qubit-cavity, and temporal mode matching via pulse shaping, which would result in an additional combined SNR improvement factor of up to seven. This would be possible without any assumed transducer performance improvements.

However, the utility of quantum networking devices depends on the trade-off between SNR and throughput, which can be quantified jointly via the quantum channel capacity (see Appendix~\ref{appendix:throughput}). To scale the quantum channel capacity of currently 2.9~Hz towards the MHz range, we need to mainly improve the throughput while keeping occupancies low. To achieve this we also identify two main drivers: (i), Materials: First tests on adding normal conductors with identical transducer geometries suggest about two orders of magnitude  
higher throughput 
mainly due to faster quasiparticle recombination~\cite{sancheznaranjo2026}. And (ii), Geometry: Optimized microwave mode localization as described in Ref.~\cite{rueda2019} is expected to yield $\approx40$ times higher $g_0$, thus reducing the required pump power and associated fiber noise, and -- assuming the low optical and microwave losses can be maintained -- provide an additional two orders of magnitude boost to the expected low noise throughput.

Only moderate throughput improvements~\cite{zhao2025,urmey2025} are needed to measure photon anti-bunching, a model-agnostic verification of the output nonclassicality~\cite{Dreau2018}, qubit-optical photon entanglement~\cite{tchebotareva2019} - a demonstration that also proves phase coherence, as well as proof of principle demonstrations of entangling remote transmon qubits through heralding and entanglement swapping~\cite{pan1998, bernien2013, krastanov2021, krutyanskiy2023a} and implementing remote gates via local operations and feed-forward~\cite{chou2018,iuliano2026}. 

Previously achieved near unity conversion efficiencies in the high cooperativity limit \cite{sahu2022,sahu2023,qiu2023} with further improved in/out-coupling efficiencies would also position the presented method as a viable candidate for comparably narrow band, transform-limited, on-demand single photon sources in the telecom band~\cite{ding2025}, even though current brightness levels (see noise-throughput tradeoff in Appendix~\ref{appendix:throughput}) would likely not be competitive with purely photonic approaches at this time.

Importantly, the demonstrated platform-independent nature of converting itinerant microwave photons -- superconducting qubits, NV centers, and electron spin qubits can all interact with MW photons -- allows for an extension of cryogenic microwave circuits via established telecom photonics. This paves the way for scalable and heterogeneous quantum networks where fast superconducting circuits and long-lived atomic qubits compute and sense together.

\section{Acknowledgements}
We thank Fritz Diorico and Onur Hosten who suggested the filter cavity design, and gave important insights about the assembly and the testing of the Fabry-Perot filter cavities. Ekatrina Fedotova and Diego A. Lancheros Naranjo worked on the filter cavity setup in the early stages of this work. Gustavo Wiederhecker and Yiewen Chu provided insights as to the origins of the observed optical noise and Nicola Carlon Zambon suggested using telecom filters to mitigate it further. This work was supported by the European Research Council under grant agreement no. 101089099 (ERC CoG cQEO), and 101248662 (ERC POC CoupledEOT), the European Union’s Horizon 2020 research and innovation program under grant agreement no. 899354 (FETopen SuperQuLAN), the European Innovation Council no. 101187231 (PathfinderOpen CIELO), and the Austrian Science Fund (FWF) no. F7105 (SFB BeyondC). J.F. and L.K. acknowledge support from the Horizon Europe Program HORIZON-CL4-2022-QUANTUM-01-SGA via Project No.~101113946 OpenSuperQPlus100. A.M. acknowledges support from the NOMIS-ISTA fellowship.

\section{Author contributions}
T.W., E.R., R.S., and S.H. conducted the measurements and analyzed the data. 
T.W., G.A., and S.H. built the microwave setup. 
E.R., R.S., and A.M. built the optical setup. 
T.W., P.F., J.A.S., and L.Q. built the electro-optic transducer. 
D.L. developed the thin film electrode fabrication. 
L.K. fabricated the superconducting qubit. 
M.Z. fabricated the JPA. 
S.H. and A.M. modelled the microwave dynamics.
T.W., E.R., S.H., A.M., and J.M.F. prepared the manuscript with input from all co-authors. J.M.F. supervised this work.

\section{Data availability}
All data and analysis code will be made available on Zenodo.

\appendix

\section{Device Parameters}
\label{appendix:device_parameters}
We list important device parameters in Table~\ref{tab:SI_device_parameters}. 
\begin{table}[h]
\begin{tabular}{ |p{6cm}|p{2cm}|  }
 \hline
 \multicolumn{2}{|c|}{\textbf{Electro-optic transducer}} \\
  \hline
 Optical resonance frequency $f_o$   & 193~THz\\
 \hline
  Optical cavity linewidth $\kappa_o/(2\pi)$ & 11.4~MHz\\
 \hline
 Optical coupling efficiency $\eta_o$ & 0.68\\
 \hline
 Electro-optic coupling rate $g_0/(2\pi)$   & 2.9~Hz\\
 \hline
 Optical mode overlap $\Lambda^{2}$ & 0.44\\
\hline
 MW resonance frequency $f_e$ &   8.9006~GHz\\
 \hline
 MW cavity linewidth $\kappa_{\textrm{eo}}/(2\pi)$ &   1.4~MHz\\
 \hline
 MW coupling efficiency $\eta_e$ &   0.44\\
 \hline
 Cooperativity $C$ &   $4.1 \times 10^{-4}$\\
 \hline
 Internal conversion efficiency $\eta_{\textrm{int}}$ &   $1.6 \times 10^{-3}$\\
 \hline
 External conversion efficiency $\eta_{\textrm{ext}}$ &   $2.2 \times 10^{-4}$\\
 \hline
Conversion bandwidth $B_c/(2\pi)$ &   $1.3~\textrm{MHz}$\\
 \hline
 \multicolumn{2}{|c|}{\textbf{Qubit-cavity}} \\
 \hline
Resonance frequency (qubit in $\ket{g}$) $f_g$ &   8.9076~GHz\\
 \hline
Dispersive shift $\chi/(2\pi)$ &   -3.5~MHz\\
 \hline
 Linewidth $\kappa/(2\pi)$ &   1.43~MHz\\
 \hline
External linewidth (qubit drive and BSB port) $\kappa_{\textrm{e1}}/(2\pi)$ &   0.1~MHz\\
 \hline
External linewidth (readout port) $\kappa_{\textrm{e2}}/(2\pi)$ &   1.0~MHz\\
 \hline
Qubit relaxation rate $T_1$ &   19~$\mu$s\\
 \hline
Qubit dephasing rate $T_{\phi}$ &   14~$\mu$s\\
 \hline
\end{tabular}
\caption{\textbf{Transducer and qubit-cavity parameters.}}
\label{tab:SI_device_parameters}
\end{table}

\section{Microwave Setup}
\label{appendix:SI_mw_setup}

Apart from room temperature signal generation and readout electronics, we place all MW components inside a dilution refrigerator (Bluefors LD-250). We utilize Eccosorb (Bluefors IR filter), lowpass (K\&L Microwave 6L250-12000/T26000-O/O6L250-00089), and bandpass filters (Keenlion 8-12GHz Band Pass Filter) as well as switches (Radiall R573423600, Radiall R570443000), directional couplers (KRYTAR 120420), circulators (LNF 8-12 GHz Single Junction Circulator), cryogenic 50~$\Omega$ terminations (Amphenol XMA 2001-6117-00-CRYO) and attenuators (Amphenol XMA 2682-6460-dB-CRYO, Amphenol XMA 2082-6340-dB-CRYO) for signal conditioning of the input signal. The output line consist of a double-junction circulator (LNF 8-12 GHz Dual Junction Circulator), a Dimer Josephson Junction Array Amplifier (JPA), and a HEMT amplifier(LNF-LNC4\_16B).

\begin{figure}
    \centering
    \includegraphics[scale=1]{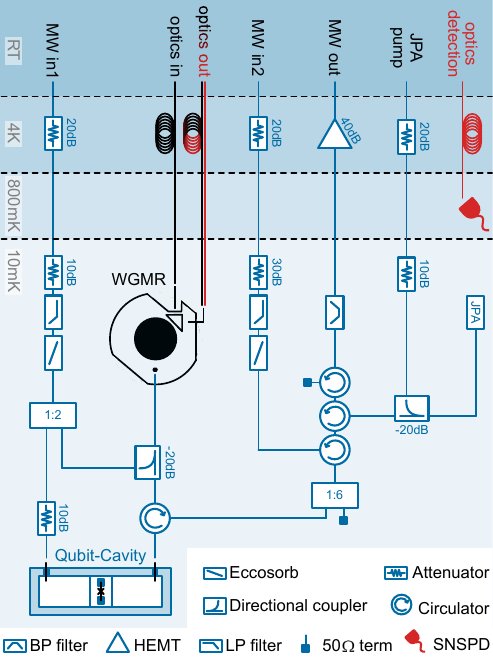}
    \caption{\textbf{Schematic of the dilution refrigerator setup.} The blue components depict cryogenic microwave equipment. Black (red) optical fibers guide the pump (signal) inside the fridge. The EO transducer is illustrated in black. The black dashed lines represent the boundaries between the blue-shaded temperature stages. All equipment, except for the fibers, is thermally anchored to the respective temperature stage. The four ports not shown on the 1-to-6 switch are connected to an unrelated sample. A SMA antenna connects the EO transducer's MW cavity to the MW setup. Band-pass (BP), low-pass (LP) and eccosorb filters reduce the noise power outside a certain range. The HEMT provides around 40~dB of gain. We write the attenuation (coupling) of the attenuators (directional couplers) next to their symbol.}
    \label{fig:SI_mw_setup}
\end{figure}

We show the cryogenic microwave setup in Fig.~\ref{fig:SI_mw_setup}. We send the qubit and BSB drives (readout pulse) from MW in1 (MW in2). We use Quantum Machines OPX+ and Octave to generate and analyze MW pulses, and to trigger the optical pulses and the SNSPD (Photon Spot) readout electronics (PicoQuant TimeHarp 260).

\section{Optical Setup}
\label{appendix:optical_setup}

\begin{figure*}[t]
    \centering
    \includegraphics[scale=1]{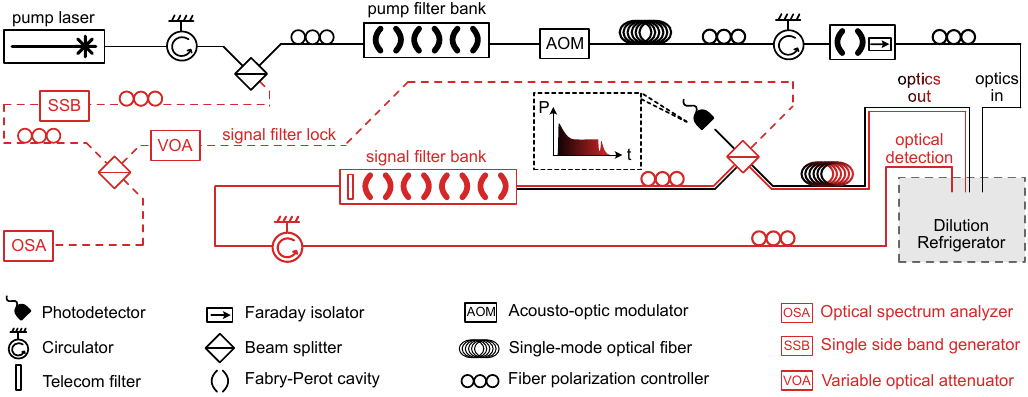}
    \caption{\textbf{Schematic of the optical setup.} The black lines represent the parts of the setup used for pump generation, pump cleaning, pulse generation, and delivery to the dilution refrigerator. The dashed red lines represent the reference tone generation and attenuation for locking the signal filters. The solid red lines represent the parts of the setup used for signal cleaning and delivery to the dilution refrigerator for detection.}
    \label{fig:SI_Setup}
\end{figure*}

We show the optical setup of this experiment in Fig.~\ref{fig:SI_Setup}. The pump is generated by a 1550~nm laser with a linewidth of $10~\mathrm{kHz}$ and a noise floor of $3~\mathrm{photons/s/Hz}$. Before being directed into the transducer, the pump tone is cleaned by three Fabry--Perot cavities aligned in series in free space. We call this setup the pump filter bank. These cavities are locked on resonance with the pump frequency. The free spectral ranges (FSRs) of the cavities are varied (staggered) by a few percent points to maintain suppression at higher-order sidebands.

\begin{figure}[t]
    \centering
    \includegraphics[scale=1]{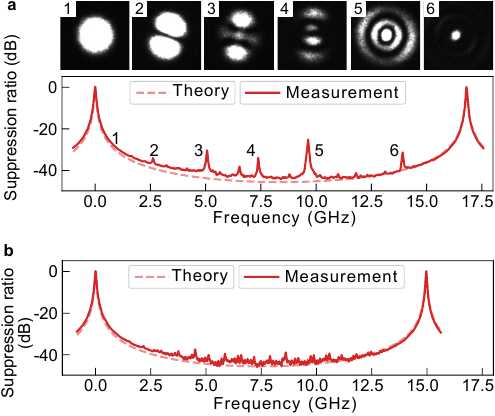}
    \caption{\textbf{Fabry--Perot filter cavity spectra.} 
    \textbf{a}, First- and higher-order transverse modes of the Fabry--Perot cavity measured by an IR camera (top images). Suppression ratio of the first filter cavity in the pump filter bank (bottom panel), with a free spectral range (FSR) of $16.8~\mathrm{GHz}$, a linewidth of $54.8~\mathrm{MHz}$, and a suppression ratio of $43.8~\mathrm{dB}$ at $8.9~\mathrm{GHz}$. The theoretical curve represents the transmission of a Fabry--Perot cavity. 
    \textbf{b}, Suppression ratio of the second Fabry--Perot cavity in the pump filter bank, with an FSR of $15.0~\mathrm{GHz}$, a linewidth of $48.8~\mathrm{MHz}$, and a suppression ratio of $43.0~\mathrm{dB}$ at $8.9~\mathrm{GHz}$.} 
    \label{fig:SI_Filter}
\end{figure}

The pump filter bank has a total on-resonance insertion loss of $2.1~\mathrm{dB}$ at the pump frequency. The measured spectra and suppression ratios of the first and second Fabry--Perot cavities in the pump filter bank are shown in Fig.~\ref{fig:SI_Filter}. The higher-order modes of the first cavity arise from slight misalignment with respect to the input Gaussian beam. Due to the technical limitations of the input beam collimator, the higher-order modes of the first cavity cannot be completely eliminated. This imperfection does not limit us as the the higher-order modes are further suppressed by a factor of $>20~\mathrm{dB}$ through coupling the Gaussian beam into a single-mode optical fiber at the end of the cavity chain.

Based on these measurements, we estimate that the pump filter bank provides an extinction ratio of approximately $130~\mathrm{dB}$ at the signal frequency. The pump pulses used for upconverting the microwave signal are generated by an acousto-optic modulator (AOM) by applying a rectangular 200~MHz RF-pulse with a duration of $200~\mathrm{ns}$. 
In order to eliminate noise photons generated at the signal frequency due to inelastic scattering of the pump pulse in the optical fiber, we use another Fabry--Perot cavity immediately before the optical input of the dilution refrigerator (DR). We also use a Faraday isolator to suppress backward-scattered photons.

To keep the conversion efficiency close to the optimal value allowed by our device, the pump light frequency is kept on resonance with the mode of the optical WGMR. To achieve this, we collect 1\% of the total reflected light off the transducer from the optical output line of the setup and monitor it with a fast photodetector in order to lock the laser frequency to the optical mode. We measure a total optical transmission loss of 4.8~dB from the input to the output of the dilution refrigerator setup. 

The remaining $99\%$ signal is cleaned by four Fabry--Perot filter cavities (signal filter bank), identical to the pump filter bank cavities, which are locked on resonance with a reference tone at a frequency of $\omega_p + 8.9~\mathrm{GHz}$. This reference tone is derived by extracting a small fraction of the pump light and upconverting it with a single-sideband modulator (SSB). The reference tone is attenuated by a variable optical attenuator (VOA) to prevent latching of the SNSPD used for locking the signal filter bank and detecting the upconverted optical signal. The SNSPD has an efficiency of 84.7\%.

The SSB is continuously on and is turned off only for $2~\mu\mathrm{s}$ during which the upconversion pump pulse is sent to the fridge. We use a telecom filter with a passband of $250~\mathrm{GHz}$ at the end of the signal filter bank, providing approximately $60~\mathrm{dB}$ suppression outside this range. The signal filter bank has a total on-resonance insertion loss of $3.6~\mathrm{dB}$ at the signal frequency while providing approximately $170~\mathrm{dB}$ suppression at the pump frequency. 
The optical reflections from the optical components in the setup are suppressed by optical circulators.

The total optical detection loss (6.7~dB) is the sum of the loss from the EO device disk-prism interface to the fridge output (2.4~dB), the transmission loss of the signal cavities (3.6~dB) and the detection inefficiency of the SNSPD (0.7~dB).

\section{Optical Noise and Upconverted Noise}
\label{appendix:optical_noise}
\begin{figure}[t]
    \centering
    \includegraphics[scale=1]{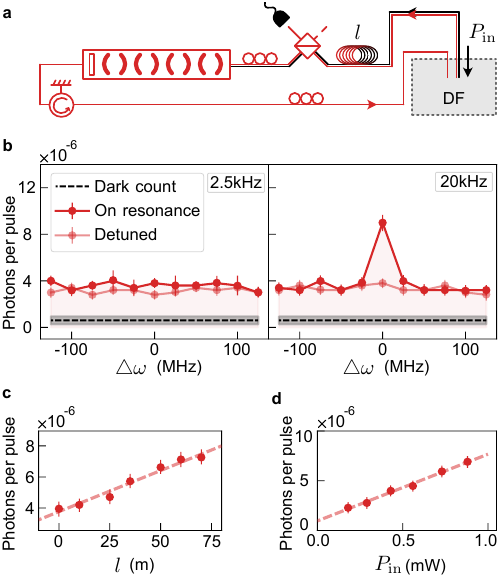}
    \caption{\textbf{Optical noise characterization.}
    \textbf{a}, Schematic of the optical setup used for optical noise characterization, indicating the optical peak power and the additional fiber lengths in the setup (the configuration is identical to that shown in Fig.~\ref{fig:SI_Setup}).
    \textbf{b}, Optical noise spectrum measured for two cases: the microwave cavity of the WGMR tuned to $8.9006~\mathrm{GHz}$ (on resonance) and detuned by $13~\mathrm{MHz}$ (detuned). Measurements are performed at two repetition rates: $2.5~\mathrm{kHz}$ (left panel) and $20~\mathrm{kHz}$ (right panel). 
    \textbf{c}, Detected optical noise as a function of the additional fiber length inserted into the optical setup at the output of the WGMR, measured with the signal lock reference tone detuned by $+100~\mathrm{MHz}$. 
    \textbf{d}, Detected optical noise as a function of the optical peak power of the input pump, measured with an added fiber length of $65~\mathrm{m}$ and with the signal lock reference tone detuned by $+100~\mathrm{MHz}$.}
    \label{fig:SI_Noise}
\end{figure}

The optical noise spectrum is measured by sending a pump pulse to the optical input line of the setup with a peak power of about $2.1~\mathrm{mW}$, which, after accounting for setup losses, corresponds to $P_{\mathrm{in}}\approx 1.2~\mathrm{mW}$ at the transducer. The output light from the signal filter bank is then detected using an SNSPD, as shown in Fig.~\ref{fig:SI_Noise}a.
To measure the frequency dependence of the noise, we detune the reference tone, used to lock the signal cavities, from $\omega_p + 8.9~\mathrm{GHz}$. We perform this measurement under two different configurations, (i) the resonance frequency of the EO MW cavity is $8.9~\mathrm{GHz}$, matching the optical FSR and enabling upconversion, and (ii) the resonance frequency is detuned from the optical FSR value by $13~\mathrm{MHz}$. We show the results of these two measurements for two trigger rate values, 2.5~kHz and 20~kHz, in Fig.~\ref{fig:SI_Noise}b.

The constant noise contribution (light red) that is independent of frequency and of trigger rate is the so-called optical noise of $(3.2\pm0.2)\times 10^{-6}$. It matches the optical noise of $(2.6\pm0.6)\times 10^{-6}$ that we extracted from the data in Fig.~\ref{fig:Noise}b within uncertainty. The optical noise consists of two main components: dark counts of the SNSPD, and broadband inelastic scattering of the pump in the optical fibers~\cite{dainese2006,kovalev2007,florez2016}. In principle, one could also observe a contribution to optical noise given by pump photons that leak through the signal cavities. The observed optical noise, however, does not follow the shape of the reflected pump. This means that the signal filter bank suppresses the reflected pump to values much lower than than dark counts. 

The dark counts are as low as $(0.6\pm0.2)\times 10^{-6}$. This makes up less than a quarter of the count rate of the optical noise of $(2.6\pm0.4)\times 10^{-6}$. The biggest contribution is due to inelastic scattering amounting to $(2.0\pm0.6)\times 10^{-6}$. To further characterize this inelastic scattering, we study its dependence on fiber length and optical peak power. We show this dependence in Fig.~\ref{fig:SI_Noise}c and d, where we observe a linear scaling that confirms the fiber and pump origin of this noise source.

Besides the optical noise, there's a trigger-dependent contribution to the noise that we measure in Fig.~\ref{fig:Noise}b. In Fig.~\ref{fig:SI_Noise}b, we see that higher trigger rates lead to a noise contribution that is visible only when the resonance frequency of the EO MW cavity equals the optical FSR. As discussed in the main text, we identify this noise contribution as thermal photons inside the EO MW cavity that get upconverted to the optical domain once the pump photons populate the optical cavity. For the 20~kHz measurements, the respective noise counts of $(9.0\pm0.4) \times 10^{-6}$ and $(9.9\pm0.4) \times 10^{-6}$ in Fig.~\ref{fig:SI_Noise}b and Fig.~\ref{fig:Noise}b nearly match within their uncertainties. At lower trigger rates, the upconverted noise is smaller. We should still see a slightly pronounced peak for 2.5~kHz in Fig.~\ref{fig:SI_Noise}b when comparing to the values in Fig.~\ref{fig:Noise}b. We assume that this discrepancy stems from fluctuations in the laser polarization which leads to lower pump population in the WGMR and therefore to lower MW noise occupation.

\section{Microwave Dynamics}
\label{appendix:SI_mw_losses_and_noise}

We use \verb +QuTiP+~\cite{johansson2013} to solve the Lindblad master equation describing the time dynamics of the qubit-resonator joint density matrix. This master equation takes the form
\begin{equation}
\label{eq: master equation}
\dot{\rho} = \frac{1}{i\hbar} [f(t) H_{\textrm{BSB}}, \rho] + \gamma_1 \mathcal{D}[b]\rho + \gamma_\phi \mathcal{D}[b^\dagger b]\rho + \kappa \mathcal{D}[a]\rho,
\end{equation}
where $\mathcal{D}[\mathcal{O}]\rho = \mathcal{O}\rho \mathcal{O}^\dagger - \{\mathcal{O}^\dagger\mathcal{O},\rho\}/2$ represents a Lindblad dissipator associated to a collapse operator $\mathcal{O}$, $[\cdot, \cdot]$ is the commutator, and $\{\cdot, \cdot\}$ is the anticommutator. The BSB Hamiltonian $H_{\textrm{BSB}} = \ket{g,0}\bra{e,1} + \ket{e,1}\bra{g,0}$ represents the two-photon transition that populates the resonator with one photon while exciting the qubit to the $\ket{e}$ state simultaneously. The time dependence $f(t)$ of the BSB drive is a rectangular pulse of 156~ns duration with strength chosen to maximize the excited state population of the qubit after the drive. In the Lindblad dissipator terms, $\gamma_1 = 1/T_1$ represents the decay rate of the qubit, $\gamma_\phi = 1/T_\phi$ the dephasing rate, and $\kappa$ the decay rate of the resonator (see Tab.~\ref{tab:SI_device_parameters} for numerical values). The operator $b = \vert g\rangle\langle e\vert + \sqrt{2}\vert e\rangle\langle f\vert$ represents the annihilation operator for the transmon, while $a$ is the standard bosonic annihilation operator for the resonator. To simulate the generation of the HP state, we initialize the joint density matrix to the state $\rho_0 = \vert +, 0 \rangle\langle 0,+\vert$ (with $\vert + \rangle = (\vert g \rangle + \vert e\rangle)/\sqrt{2}$), let it evolve under Eq.~\eqref{eq: master equation}, and extract the resonator's Fock state $\vert 1\rangle$ population $P_1(t) = \mathrm{Tr}(\vert 1\rangle\langle 1\vert \rho(t))$. The population $P_1(t)$ corresponds then to the time envelope of the generated HP state while it propagates through the output waveguide towards the EO resonator. We can understand this fact under an input-output setting~\cite{gardiner1985}, where the dissipator $\mathcal{D}[a]$ represents the decay of the resonator's state towards its output modes.

The grey solid line in Fig.~\ref{fig:FockStates}a is the simulated population $P_1(t)$ rescaled in amplitude to match the experimental data. When the EO resonator is off-resonant, the generated pulse reflects without any change in its time envelope. When the EO resonator is resonant, we need to account for the fraction of the microwave population that gets absorbed by the EO resonator, and later re-emitted. To do so, we model the resonator with a frequency domain transfer function $H(\omega) = \eta_e \kappa_{eo} / (i\omega + \kappa_{eo}/2)$ (see Tab.~\ref{tab:SI_device_parameters} for numerical values). Then, we derive the black solid line in Fig.~\ref{fig:FockStates}a by Fourier transforming the off-resonant reflection envelope $\mathcal{F}(S_{\textrm{MW}})$, multiplying it with $(1-H(\omega))$, and by applying the inverse Fourier transform to the result.

We also use $H(\omega)$ to calculate the single photon occupation during the conversion process $N_{\textrm{SP}}$. $N_{\textrm{SP}}$ assumes that one itinerant photon is present at the input of the EO MW cavity. In this case, we multiply $H(\omega)$ with $\mathcal{F}(S_{\textrm{MW}})$, inverse Fourier transform it and convolve the result with a 200~ns rectangular pulse, representing the optical pump. This yields $N_{\textrm{SP}} \approx 0.26$ which is a factor $\approx2$ smaller than expected for a coherent state. We choose an optical pulse of 200~ns duration since $N_{\textrm{SP}}$ does not decrease significantly from its optimum at $1/B_c \approx 122~\textrm{ns}$  (0.27) and the longer pulse results in higher count rates.

We calculate the two curves describing models in Fig.~\ref{fig:FockStates} using HP data because they contain phase information for the two field quadratures I and Q. We scale the HP data with a multiplication factor of 5.57 and add a MW noise offset, which is the sum of the input referred added noise (1.84) and vacuum noise (0.5) multiplied with the detection bandwidth of 50~MHz, to fit the SP data. The theoretical scaling factor is \[\frac{\abs{\ev{\ha}}^2}{\ev{\had \ha}} = \frac{1}{4}\] with $\abs{\ev{\ha}}$ of HP being 0.5 and $\ev{\had \ha}$ of SP being 1. The theory scaling factor ($10\times \log_{10} 4$) and the fit scaling factor ($10\times \log_{10} 5.57$) differ by 1.5~dB. The reference point for the input referred added noise is a 50~$\Omega$ termination with known temperature at the 1-to-6 switch.

\section{Microwave SNR and Loss}
\label{appendix:SI_mw_losses_and_SNR}
We define $\alpha$ representing the loss between inside the qubit-cavity and the input of the JPA. To estimate $\alpha$, we first determine the loss from the 1-to-6 switch to the qubit-cavity, then to the EO transducer and back to the 1-to-6 switch. We do this by first connecting the input of the switch to the port which goes to the qubit-cavity. In this configuration, we measure the loss between MW in2 and MW out (see Fig.~\ref{fig:SI_mw_setup}). Then we put the switch to a state in which it reflects all incoming signals and measure again. The difference between these two measurements shows a 4~dB lower loss when the switch is in the reflective state. We use the data sheet values for insertion and transfer losses of our used components to distinguish the individual components' loss contributions. They are listed in TABLE~\ref{tab:SI_mw_losses}. Using these values we estimate the losses between the qubit-cavity and the EO transducer $L_{\textrm{eo,qc}} = 1.6~\textrm{dB}$, the transducer and the switch $L_{\textrm{sw,eo}} = 1.6~\textrm{dB}$ and between the switch and the JPA $L_{\textrm{JPA,sw}} = 1~\textrm{dB}$. The predicted $\alpha$ is 0.27.

Combing the outcoupling efficiency of the qubit-cavity ($\kappa_{\textrm{e2}} / \kappa \approx 0.7$) and $L_{\textrm{eo,qc}}$ results in $\approx0.5$ photons reaching the input port of the EO MW cavity when a single photon is generated in the qubit-cavity. The effective MW single photon occupation in the transducer is \[N_{\textrm{cav,SP}} = 0.5\,N_{\textrm{SP}} \approx 0.13.\]

If the EO MW cavity is detuned from the qubit-cavity, the loss between the qubit-cavity and the switch is $L_{\textrm{sw,qc}} = L_{\textrm{eo,qc}} + L_{\textrm{sw,eo}} = 3.2~\textrm{dB}$. If we add the qubit-cavity outcoupling efficiency to this, we predict that 0.334 photons arrive at the switch. If we integrate the off resonance curve in Fig.~\ref{fig:FockStates}b over 800~ns and remove the integrated noise, we calculate 0.337 photons at the $50~\Omega$ reference at the switch. So, the predicted and measured values for the loss match well. We choose 800~ns since it covers nearly the full shape of the photon. At longer integration times, the integrated noise has a bigger impact. Dividing the integrated signal by the integrated noise yields an SNR of 0.0038. We use a 50~MHz IF bandwidth (IFBW) for this measurement.

The single photon Rabi measurement in Fig.~\ref{fig:RabiMeasurements}c uses a MMF with a duration of 800~ns. Dividing the amplitude of the Rabi oscillation (0.16) value by the noise offset (input referred added noise + vacuum = 2.34), yields an SNR of $\approx 0.069$. Due to the MMF, the noise is measured at a equivalent noise band width of $\textrm{ENBW} = 2.03~\textrm{MHz}$~\cite{harris1978}. If we account for the different bandwidths of the measurements in Fig.~\ref{fig:FockStates}b and Fig.~\ref{fig:RabiMeasurements}c, we can compare their respective SNRs. Scaling the SNR of the Rabi measurement by ENBW/IFBW, results in an SNR of $\approx0.0028$. This differs from the predicted SNR (0.0038) by 1.2~dB.

\begin{table}
\begin{tabular}{ |p{6cm}|p{2cm}|  }
  \hline
 Transfer loss copper cable & 1~dB/m\\
 \hline
  Insertion loss directional coupler & 0.5~dB\\
 \hline
 Insertion loss single junction circulator & 0.2~dB\\
 \hline
 Insertion loss double junction circulator & 0.3~dB\\
 \hline
\end{tabular}
\caption{\textbf{Microwave insertion and transfer losses}}
\label{tab:SI_mw_losses}
\end{table}

\section{Transducer Throughput}
\label{appendix:throughput}

\begin{figure}[t]
    \centering
    \includegraphics[width=\linewidth]{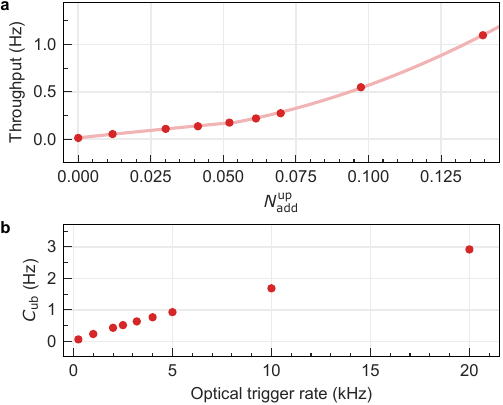}
    \caption{\textbf{Throughput and Quantum channel capacity.} \textbf{a}, The faint red line represents two fits to the data. Up to $N^{\textrm{up}}_{\textrm{add}} \approx0.05$, the fit is linear $\propto {N^{\textrm{up}}_{\textrm{add}}}^{0.94\pm0.004}$. From there, it is quadratic $\propto{N^{\textrm{up}}_{\textrm{add}}}^{2.03\pm0.003}$. The referred added input noise values of the data correspond to the optical trigger rates of the data in \textbf{b}. \textbf{b}, The red dots represent the upper bound of the quantum channel capacity $C_{\textrm{ub}}$.}
    \label{fig:SI_throughput}
\end{figure}

To compare between transducers operating under different conditions with different operating principles, one needs a transduction-detail agnostic figure of merit. Throughput is one such figure of merit that applies to transducers operating in pulsed as well as continuous operation~\cite{urmey2025}. 
It is defined as the product of the conversion bandwidth ($B_c$), the duty cycle ($D$) and the conversion efficiency ($\eta_{\textrm{ext}}$), \[\Theta = B_c \times D \times \eta_{\textrm{ext}}.\] This formula treats the transducer as a black box. It gives the rate of generating upconverted photons assuming we continuously apply a signal with one microwave photon per conversion bandwidth at the input. We note that the $\eta_\textrm{{ext}}$ we use considers a coherent input~\cite{hease2020a}.

We plot the throughput that our transducer achieves as a function of the added conversion noise referred to the MW input \[N^{\textrm{up}}_{\textrm{add}} = \frac{N_e}{\eta_e}\] in Fig.~\ref{fig:SI_throughput}a. The occupation of the microwave mode $N_e$ is the sole contribution to the referred input noise. We do not regard the optical noise because it only depends on the optical setup and not on the transducer.

The throughput data fits well to power laws whose exponents match the inverse of the exponents of the power law fits to the MW mode occupation in Fig.~\ref{fig:Noise}b. This demonstrates that we are only limited by the upconverted thermal noise from the transducer MW mode.

As indicated in the main text, we cannot quantify the mode occupation at a trigger rate of 250~Hz. Therefore, we set $N^{\textrm{up}}_{\textrm{add}}$ to zero and use the added upconversion noise $N^{\textrm{up}}_{\textrm{add}} = 0.0118\pm0.0001 \approx 0.012$ at 1~kHz as the lowest that the transducer can reach. We observe that, at higher $N^{\textrm{up}}_{\textrm{add}}$, the throughput increases much faster than the added noise meaning that just by increasing the trigger rate, more qubits can be transmitted using our transducer without adding too much noise.

As our device is quantum-enabled for the whole range of trigger rates $N^{\textrm{up}}_{\textrm{add}} < 1$ and $\eta_{\textrm{ext}} \ll 1$, we can calculate an upper bound to the quantum channel capacity over our transducer's bandwidth~\cite{pirandola2017, wang2022, urmey2025}

\begin{equation}
C_{\textrm{ub}} = \frac{\pi\Theta}{\ln{2}}(1 - N^{\textrm{up}}_{\textrm{add}} + N^{\textrm{up}}_{\textrm{add}}\ln{N^{\textrm{up}}_{\textrm{add}}}).
\end{equation}

For $N^{\textrm{up}}_{\textrm{add}} \to 0$, this turns into 

\begin{equation}
C_{\textrm{ub,0}} = \frac{\pi\Theta}{\ln{2}}.
\end{equation}

We show $C_{\textrm{ub}}$ plotted against the optical trigger rate in Fig.~\ref{fig:SI_throughput}b. For the point at the lowest trigger rate (250~Hz), we plot $C_{\textrm{ub,0}}$. At a trigger rate of 20kHz, we can reach a maximum $C_{\textrm{ub}}\approx$~2.91~Hz, meaning that we can distribute a maximum of about three ebits of entanglement using this transducer in future experiments~\cite{bennett1997,pirandola2017}.

\section{SNR Requirements for Future Experiments}
\label{appendix:Bell_fidelity}

\begin{figure}[t]
    \centering
    \includegraphics[width=\linewidth]{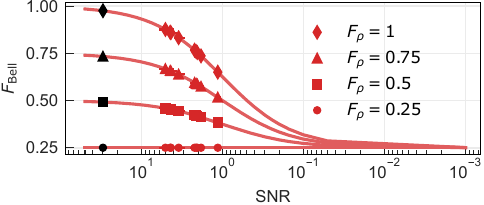}
    \caption{\textbf{Bell state fidelity vs. SNR.} The red markers represent an upper bound for the Bell state fidelity for measured SNR values (see Fig.~\ref{fig:Noise}a) with varying pre-detection fidelities. The faint red lines connecting the markers are theory lines for the respctive $F_{\rho}$. The black markers indicate varying $F_{\rho}$ for an SNR = 30, a realistic projection for future experiments. The errors on the measured values are one standard deviation.}
    \label{fig:SI_Bell_state}
\end{figure}

In this section, we comment on the SNR requirements for future transduction experiments and comment on what experiments and performance our current transducer can allow. The next major goal for transducers is to show entanglement between a superconducting qubit and an optical photon, which can be later used to distribute entanglement. We make estimates on what performance is achievable in an upconversion experiment where a loss-robust state encoded in a travelling MW state is upconverted. Such experiments have been performed in NV centers using a time-bin encoded photonic qubit entangled with a MW solid-state qubit~\cite{tchebotareva2019}.

To estimate the performance of entangled state upconversion by our transducer, we consider a very simple model where we assume that the MW photon state being upconverted is constructed using single-photon Fock states such that the loss events can be heralded away up to extra noise photons. We also assume that when a conversion event happens, it is perfect i.e. the state from the MW domain maps on to the optical photons without any decoherence. Under these conditions, whenever an optical photon is detected, it can come from one of two sources: the upconverted MW state with probability $P_\textrm{success}$, or noise otherwise. As the noise photons have no correlation with the qubit state, we consider that the total state maps to a maximally mixed state $\mathbb{I}$ when a noise event happens. In effect, the state $\rho_i$ after conversion maps to a $\rho_f$,
\begin{equation}
    \rho_f = P_\textrm{success}\cdot \rho_i + (1-P_\textrm{success})\cdot \mathbb{I},
\end{equation}
with \begin{equation}
    P_\textrm{success} = \frac{\textrm{SNR}}{\textrm{SNR} + 1}.
\end{equation}

To quantify the effect of this noise addition on the quantumness of the input state, we can consider the metric of the fidelity with a Bell state $F = \ev{\rho}{\Psi^+}$, as it provides a sharp threshold at a fidelity of $0.5$ to distinguish between states that can be classically described, and those that need a quantum formalism~\cite{sackett2000}. As this metric is also a linear function, we have the Bell fidelity $F_\textrm{Bell}$ of the final state in terms of the Bell fidelity $F_\rho$ of the initial state as
\begin{equation}
    F_\textrm{Bell} = P_\textrm{success} \cdot F_\rho + (1-P_\textrm{success})\cdot (1/4).
\end{equation}
We show the dependence of this Bell state fidelity on SNR and $F_{\rho}$ in Fig.~\ref{fig:SI_Bell_state} with the red markers indicating the achievable $F_\textrm{Bell}$ for the current SNR values, and the black markers indicating the achievable $F_\textrm{Bell}$ after the expected improvements in the setup increasing the maximum SNR to 30. For a perfect input Bell state, we note that the current maximum SNR of $5.1$ gives a $F_\textrm{  Bell} = 0.877$ indicating that our current transducer is good enough for the future experiments and would not limit the fidelity of the upconverted Bell state. We note that for the case of a time-bin entangled scheme~\cite{tchebotareva2019}, the measured SNR from Fig.~\ref{fig:Noise}a halves because the process incorporates one MW photon and two optical pump pulses.
\begin{figure}[t]
    \centering
    \includegraphics[scale=1]{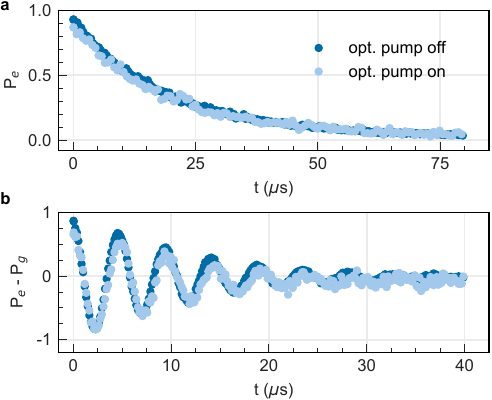}
    \caption{\textbf{Dependance of qubit parameters on optical pump.} \textbf{a} (\textbf{b}), shows two measurements of the qubit relaxation time $T_1$ (qubit coherence time $T_2$)~\cite{blais2021}. The light blue markers represent measurements when optical pump pulses arrive at the device right before the qubit readout pulse. The dark blue markers represent measurements when the optical pump is off. P$_e$ (P$_g$) is the excited (ground) state population of the qubit.}
    \label{fig:SI_optical_impact}
\end{figure}

\section{Optical Impact on Qubit}
\label{appendix:SI_optical_impact}
Optical radiation is known to have a destructive effect on superconductivity and therefore degrades the performance of superconducting qubits on exposure. To avoid the deleterious effects of optical radiation from the transducer on our superconducting qubit, we shield the qubit-cavity from the optical pump by utilizing a modular approach. Since the qubit is not on the same chip as the transducer, we can employ $\mu$-metal and aluminized \verb +Mylar+ shields around it. The modular approach also allows us to include a circulator between the qubit-cavity and the EO transducer's MW cavity. This shields the qubit-cavity from generated MW noise leaving the transducer (see Fig.~\ref{fig:SI_mw_setup}). 

To demonstrate that the qubit is insensitive to the optical pump used for transduction, we separately measure its relaxation time $T_1$ and coherence time $T_2$ with the optical pump being turned on or off. In these measurements, we initialize the joint qubit-cavity state to $\ket{e,0}$, which makes the qubit resonator resonant with the EO MW cavity. We show measurements of $T_1$ and $T_2$ in Fig.~\ref{fig:SI_optical_impact}~\cite{arnold2025}.

To measure $T_1$, we drive the qubit to the excited state with a $\pi$-pulse and measure its excited state population after a time $t$. When we measure the $T_1$ in the presence of the optical pump, we time the pump pulse to arrive right before the MW readout pulse. This means that during the whole measurement the optical pulse either overlaps with the initialization pulse (for small delay time $t$) or arrives between the qubit and the readout pulse (for larger delay times $t$). We compare this measurement to a measurement where we do not send an optical pump pulse. The fitted values for $T_1$, 19.0~$\pm$~0.4 with the optical pulse being on and 19.03~$\pm$~0.13 for it being off, agree well.

To measure $T_2$, instead, we perform a Ramsay measurement where we send the optical pump pulse right before the second $\pi/2$ pulse. If the optical pulse is on (off), the measurement yields a $T_2$ of 11.32~$\pm$~0.33 (11.63~$\pm$~0.14).

In conclusion, we see that the optical pump does not degrade the qubit parameters. This is especially important for experiments containing pulse sequences with multiple optical pump pulses, like heralded entanglement schemes~\cite{tchebotareva2019}.

\end{document}